\documentclass[prl,twocolumn,showpacs,preprintnumbers,amsmath,amssymb,floatfix,superscriptaddress]{revtex4}

\usepackage{graphicx}
\usepackage{dcolumn}
\usepackage{bm}
\newcommand{\rar}{\rightarrow}

\begin{document}
\title{Tracing magnetism and pairing in FeTe-based systems}

\author{K.~Palandage}
\email{kalum.palandage@trincoll.edu}
\affiliation{Department of Physics, Trinity College,
 Hartford, CT 06106 }

\author{K.~Fang, G.~W.~Fernando}
\email{fernando@phys.uconn.edu}
\affiliation{Department of Physics, University of Connecticut, Storrs, CT 06269 }

\author{A.~N.~Kocharian}
\affiliation{Department of Physics and Astronomy, California State University,
Los Angeles, CA 90032}



\pagestyle{plain}
\setcounter{page}{1}
\pagenumbering{arabic}

\begin{abstract}
 In order to examine the interplay between magnetism and superconductivity,  we monitor the
 non-superconducting chalcogenide FeTe and follow its
 transitions under insertion of oxygen, doping with Se and vacancies of Fe using spin-polarized band structure methods (LSDA with GGA) starting from the
 collinear
 and bicollinear   magnetic arrangements.
 We use a supercell of Fe$_8$Te$_8$  as our starting point so that it can capture local changes in magnetic moments.
 The calculated values of magnetic
 moments agree well with available experimental data while oxygen insertions lead to
 significant changes in the bicollinear or collinear magnetic moments.
 The total energies of these systems indicate that the collinear-derived structure is the more favorable one prior to
 a possible superconducting transition. Using a 8-site Betts-cluster-based lattice  and the Hubbard model, we show why this structure
 favors electron or hole pairing and provides clues to a common understanding of charge and spin pairing in the cuprates, pnictides and chalcogenides.

\end{abstract}
\maketitle

\setlength{\textheight}{22cm}

     High temperature superconductors discovered in the 1980s consisted  of  copper-oxide-based, layered
materials. In such cuprates, it is believed that doping, away from half filling, of a Mott-insulator leads to superconductivity
 although there is still no general agreement on a specific mechanism.  Recently discovered
     superconductivity in Fe-based compounds, which are either pnictides or chalcogenides~\cite{kami, fong}, has opened up  an extremely rich and active area of basic research.
    Although the transition temperatures of the Fe-chalcogenide superconductors are among the lowest of the recently discovered compounds,
they possess rather simple layered structures and fascinating antiferromagnetic or spin density wave (SDW) states.
     Superconductivity in the Fe-chalcogenides, having the so-called (11) structure, was first reported in 2008 (Refs.~\cite{ fong, fang} ) in Fe$_{1+\delta}$Se and
 Fe(Se$_{1-x}$Te$_{x})_{0.82}$.
This discovery led to a substantial increase in research efforts focused on simple, layered Fe compounds, containing chalcogenides such as S, Se, and Te.
    FeTe, which is a metallic antiferromagnet, has a tetragonal structure and has shown properties uniquely different from some of the other pnictides and chalcogenides.
 For example, Fe$_{1+\delta}$Te is not superconducting; instead, it shows magnetic and structural transitions at 65 K~\cite{katayama}.
 In addition, it is said to have the so-called ``double stripe" antiferromagnetic order instead of the ``single stripe" with ordering vector
$(0.5,0.5)$~\cite{lima}.
 Nevertheless, FeTe doped with Se
 was found to be superconducting (Refs.~\cite{xu, kami}).
Oxygen insertions in FeTe (films) have also given rise to superconductivity (Refs.~\cite{yuefeng, si}).
There are claims that  the Fe-chalcogenides do not exhibit the same nesting feature found in the pnictides and that they possess
a different local SDW magnetic order which survives even in the highest T$_c$ samples~\cite{bao}.

 The present work is focused on understanding such changes while
attempting to find common behavior, especially en route to superconductivity, in these chalcogenide (and possibly pnictide and cuprate) systems.
 We note that there have been numerous band structure calculations carried out recently on related systems focusing on nesting features of
the relevant Fermi surface, susceptibilities among other things (see Ref.~\cite{moon} and references therein); however, to our knowledge,
 these studies have not focused on
systematic changes in the resulting moments and/or total energy as functions of various modulations en route to superconductivity.

\begin{figure} 
\begin{center}
\includegraphics*[width=21pc]{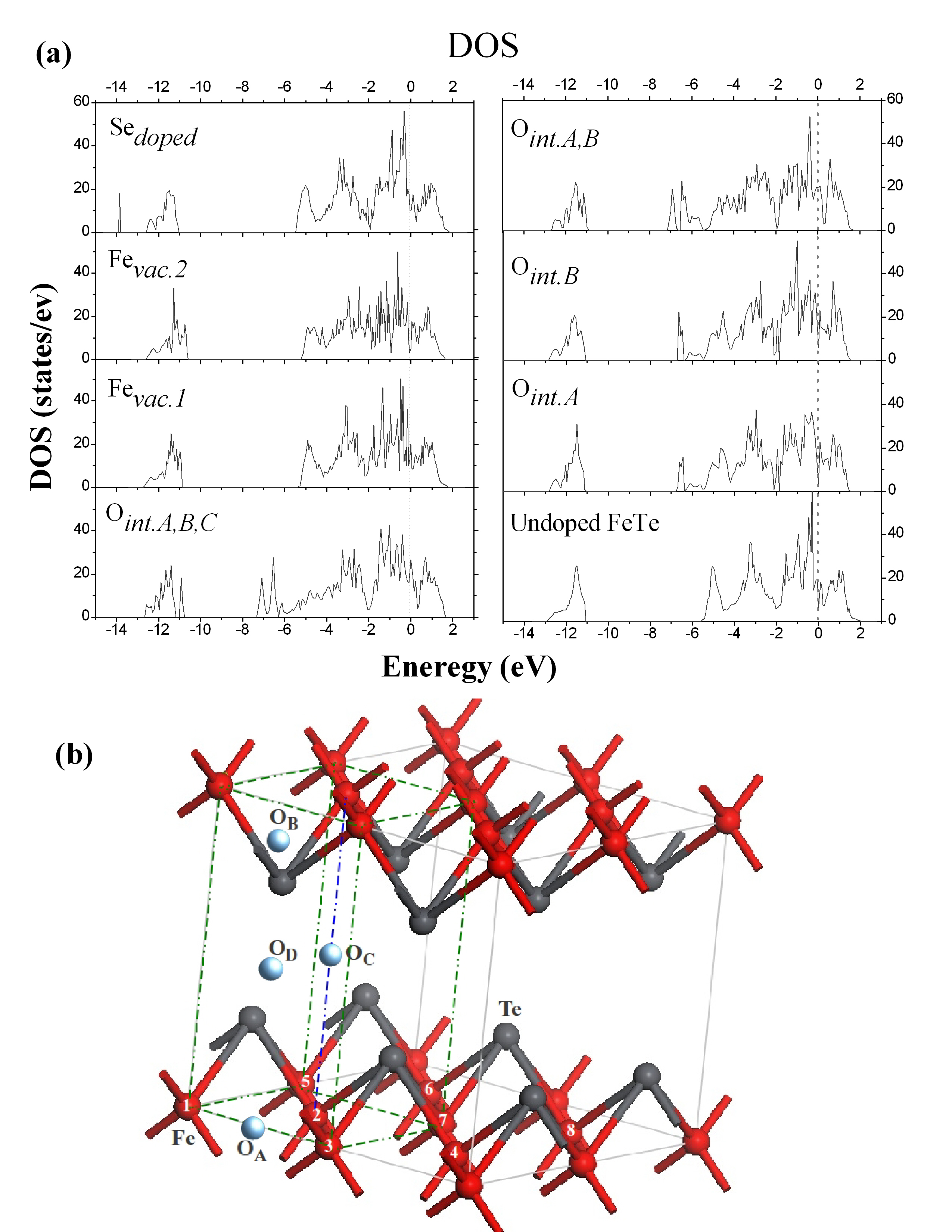}
\end{center}
\caption{Projected FeTe density of states due to various modulations and the supercell studied here;  Fe atoms are located on the  plane
 shown and possible oxygen interstitial sites are identified as A, B, C and D; Te atoms are gray spheres in between the planes of Fe.}
\label{fig:dos}
\end{figure}

\subsection{Band structure}

  Our band structure calculations are based on the
  (spin-polarized) density functional theory (DFT)~\cite{HOHENBERG,KOHN}, combined with the generalized gradient approximation (GGA)~\cite{perdew},
 as implemented in the VASP package~\cite{VASP1,VASP2,VASP3}.
The interaction between the electrons and atomic cores is described by projector augmented wave (PAW)
pseudopotentials. The wave functions are expanded using plane waves with a cut of energy of 400 eV. Brillouin Zone
integrations are carried out with a $4\times4\times4$ Monkhorst-Pack~\cite{MONKH} grid of k-points while a denser
 $8\times8\times8$ k-point mesh was used for the density of states (DOS) calculations.The lowest energy structure was determined
 using Broyden-Fletcher-Goldfarb-Shanno (BFGS)~\cite{BFGS} based algorithms. For the  results reported in this work,
 all the atoms are allowed to relax based on Hellman-Feynman forces
 and optimization is carried out until typical forces on the atoms are around $0.01\ eV/\AA$ or less.

 FeTe has a tetragonal structure with space group P4/nmm, where a square lattice of Fe  is tetrahedrally coordinated with Te ions.
 Experimental lattice parameters of the pure FeTe are $a = b = 3.8248 \ \AA$ and $c = 6.2910 \ \AA$; these were used as the starting lattice
 parameters for the $2\times2\times1$ supercell.
 This 16-atom tetragonal supercell, consists of 8 Fe and 8 Te atoms (see Fig.~\ref{fig:dos}(b)) with
  Fe atoms  located in a plane with Te atoms sitting above (and below) the plane.
  The reason for choosing a large supercell is to examine possible
 nontrivial SDW-type behavior; i.e., to allow moments to (vary and) align among themselves as necessitated by the variational principle in total energy.
 A smaller cell will not have such freedom. Following the (non-spin-polarized) work of Ref.~\cite{yuefeng},
 the interstitial sites for oxygen insertion (A, B, C, D in Fig.~\ref{fig:dos}(b)) were selected by comparing the total energies of several possible
 interstitial sites containing oxygen in the unit cell. Although our spin-polarized work shows some differences (related to sites C and D) with the above unpolarized work,
 sites A and B (for oxygen interstitials) appear to be the most energetically favorable, as determined there.

 The other modulations studied in this work are; (a) Fe interstitials, (b) Fe vacancies and (c) Se dopants.
 It is experimentally known that, except in case (a), these changes to pure FeTe, carried out with sufficient care, are likely to induce superconductivity.
 We note that without the GGA, the calculated moments turn out be rather small and (quantitative) conclusions drawn from pure LDA (or LSDA)
 work alone may be questionable.

\subsection{Magnetic Moments}

  The magnetic moments in pure FeTe and its various modulations were  calculated starting from two different (AFM) magnetic structures,
 collinear and bicollinear (see Fig.~\ref{fig:moments}).
Note that in the collinear structure, ${\vec k}= (0.5, 0.5)$ (with respect to a 2-Fe unit cell) antiferromagnetic order is present
 while the bicollinear structure has (0.5, 0.0) order.
The pure compound FeTe, in both magnetic structures, carry Fe moments which are around 2.0 $\mu_B$, in agreement with recent experiments.
This fact alone points to the relevance and importance of multi-orbital effects at the Fermi surface since otherwise the moments cannot be this high.
Our density of states plots (Fig.~\ref{fig:dos}(a)) identify these as originating from d-orbitals, which is not a surprise.
These moments are found to be severely affected by the insertion of two oxygen atoms into interstitial sites A and B (see Table~\ref{table1} and \ref{table2});
Four of the eight Fe atoms show moments that are around 1.2 $\mu_B$ or less while the total charge in each atom undergoes minimal (less that one percent) changes.
In density functional calculations based on various approximations such as the LDA or GGA, such changes are not unusual. Hence, one
cannot make specific statements about changes in the valence (such as Fe$^{2+} \rar $ Fe$^{3+}$) using these results alone.
However, when the total valence charges of Te and O are compared (at the same radii), there is a clear difference. Oxygen has more charge while being more
electronegative and hence, its insertion is likely to create holes in the Fe atoms.

\begin{figure} 
\begin{center}
\includegraphics*[width=20pc]{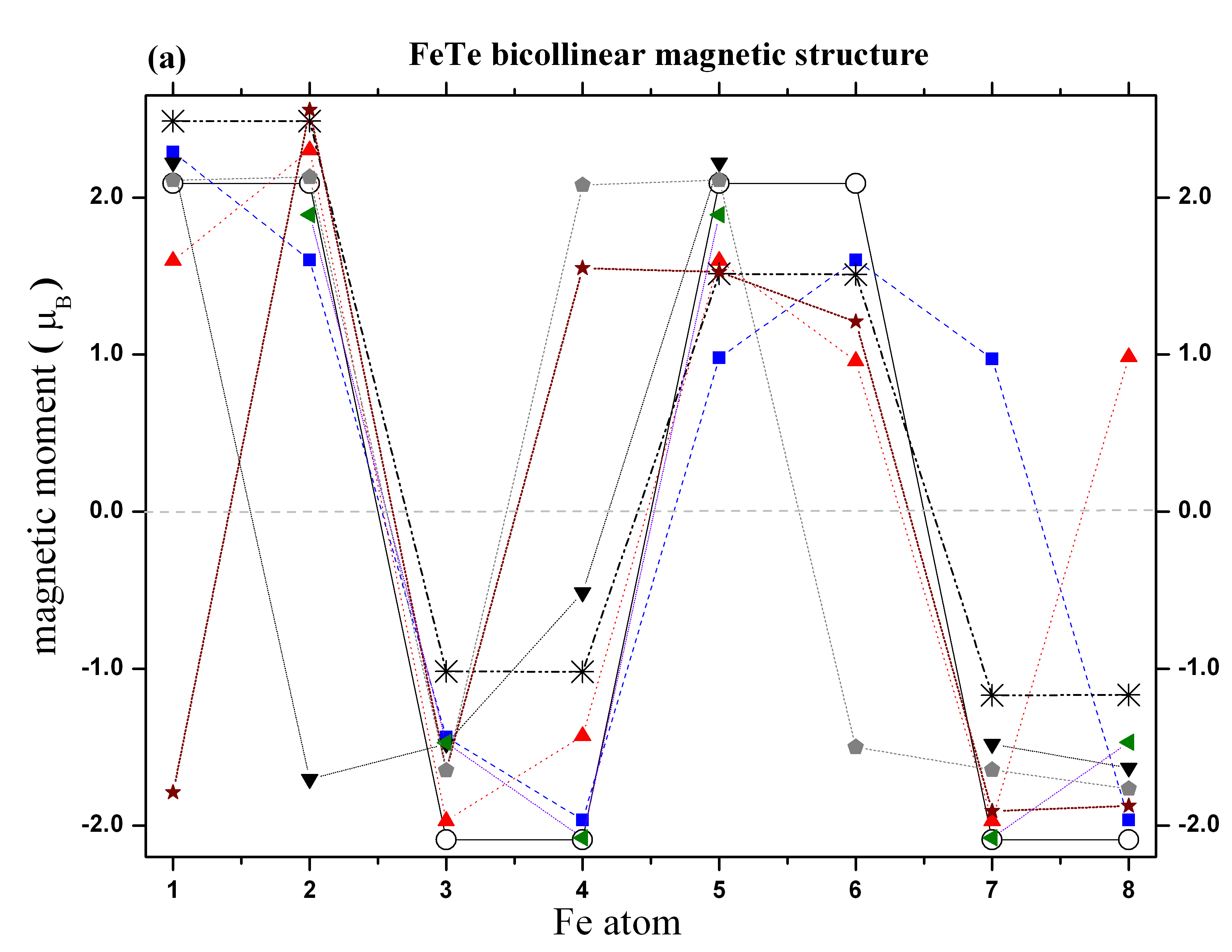}
\includegraphics*[width=20pc]{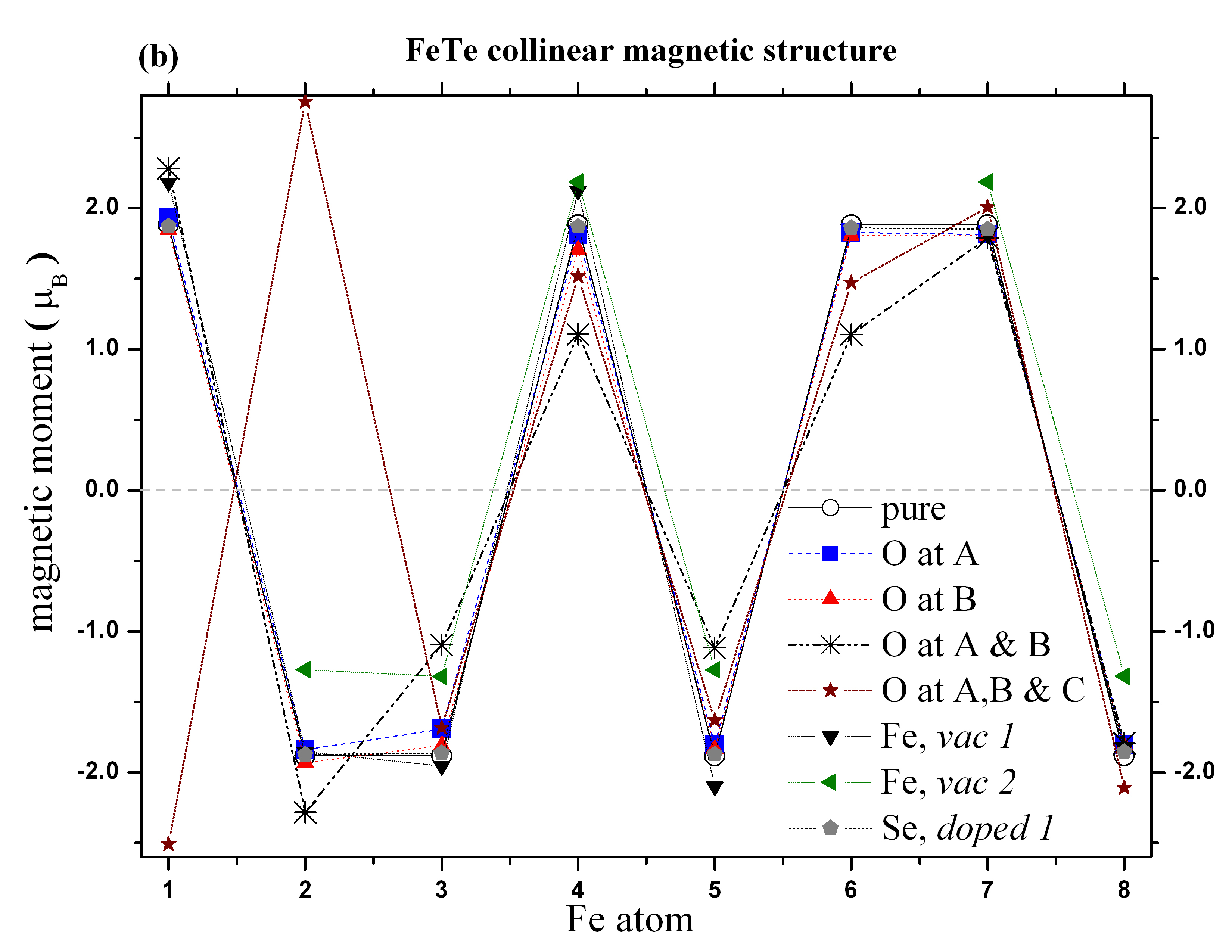}
\end{center}
\caption{Damped antiferromagnetic or SDW arrangements as calculated for bicollinear (top) and collinear (bottom) Fe arrangements. Atom labels are as in Fig.~\ref{fig:dos}.}
\label{fig:sdw}
\end{figure}

 Insertion of Fe into  A and B interstitial sites of FeTe
does not appear to change the moments as drastically. In fact, while the moments of the Fe atoms in FeTe are almost unchanged, showing long-ranged
antiferromagnetic order,
 the inserted Fe atom carries a moment of about 2.46 $\mu_B$, close to the value reported by some experiments~\cite{mart}.
The variationally determined total energies of these two, collinear and bicollinear, structures show a trend that supports experimental
 findings for FeTe and Se-doped FeTe. With the pure system
exhibiting collinear (static)  antiferromagnetic (AFM) ground state,
 we have examined the iron or oxygen inserted systems (Fe or oxygen at A and B sites, as shown in Fig.~\ref{fig:dos} and Tables~\ref{table1} and \ref{table2}
 in addition to other modulations.
Interestingly, the ground state magnetic structure of FeTe-Fe appears to depend on the extra Fe content.
As reported in Ref.~\cite{xu}, in Fe$_{1+\delta}$Te, the favored magnetic structure is the bicollinear one as our calculations indicate for
 $\delta$=0.125. For $\delta=0$, the situation is reversed and the difference in total energy between the two phases with $\delta=0$ and $\delta=0.125$
 is less than 0.2 eV (per supercell). The two magnetic states
appear to be competing to be the ground state, with total energies that are quite close to one another.

\begin{figure} 
\begin{center}
\includegraphics*[width=21pc]{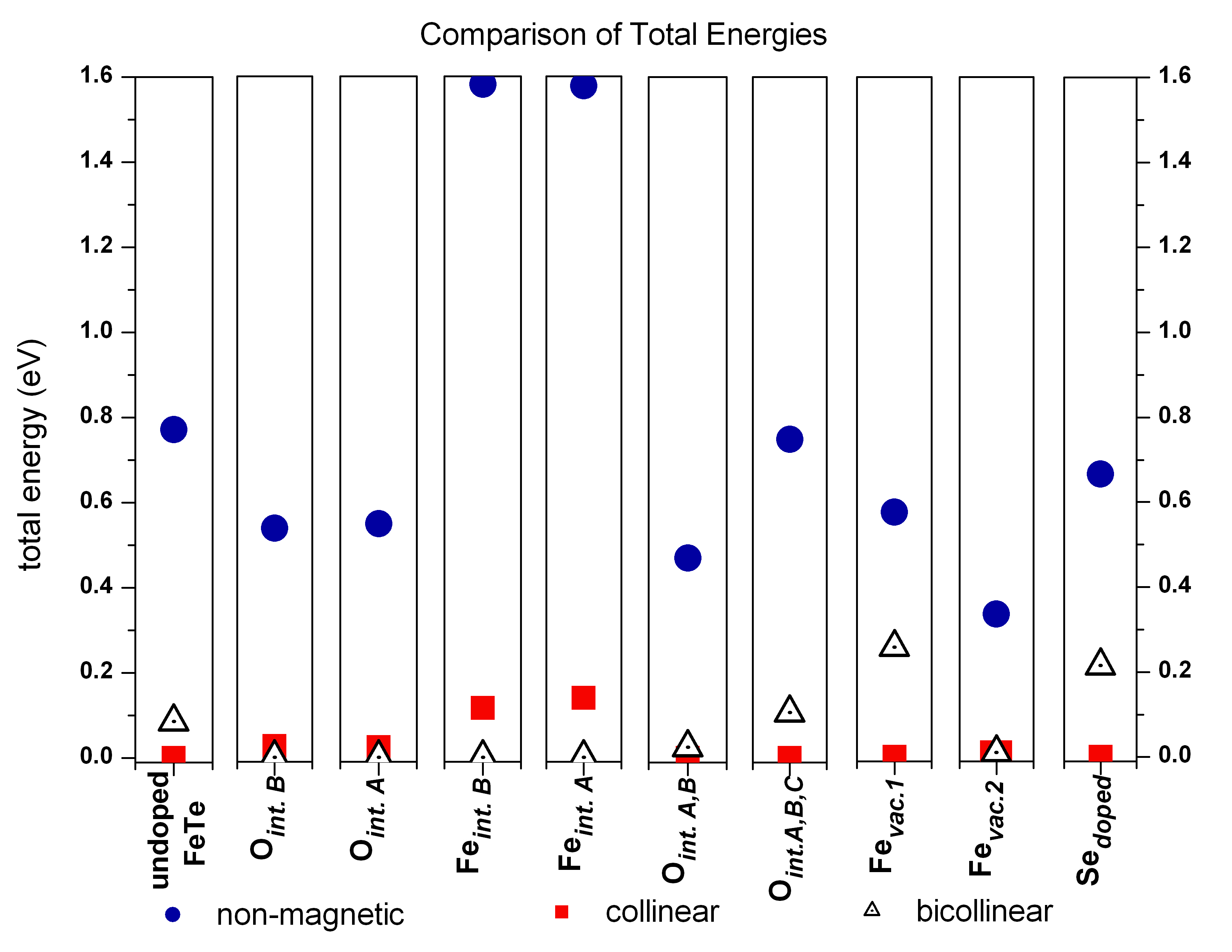}
\end{center}
\caption{Comparisons of total energies of Fe-Te magnetic structures with various modulations.
 Note that the collinear structure is the more favorable one prior to going into a possible superconducting state.}
 \label{fig:tot_energy}
\end{figure}

 However, with oxygen insertion, this situation
changes noticeably; with oxygen occupying either site A or site B, the bicollinear arrangement becomes more stable with some rearrangements in the
(magnitude and direction of) magnetic moments.
However,
 with oxygen occupying both A and B sites, the collinear structure (with reduced/different moments) becomes the more stable one.
We emphasize that this is an experimentally observed trend (Ref. ~\cite{xu}).
There is clearly no true antiferromagnetism here; instead, we see some Fe moments reduced by as much as 30-50\% while others show smaller changes.
We will label such magnetic states as damped SDW states. Our unit cell contains 8 Fe atoms  and hence it is able to show the damping/oscillations better than what would
be seen in a small (say Fe$_2$Te$_2$) cell;  a larger supercell with more Fe atoms is likely to show the damping or oscillatory effects even better.
In general, experiments show that static magnetism hinders superconductivity  while spin fluctuations, which increase
dramatically as the superconducting transition temperature T$_c$ is approached, help or act as a precursor to it in the cuprates, pnictides and chalcogenides.
 Another point to note is the  reduction in some Fe-site moments in the Fe$_{8-\alpha}$Te$_8$ calculation with two vacancies in the collinear structure;
 however, in every case some notable local magnetic order is still present.

\begin{table}[t]
\begin{center}
\renewcommand{\arraystretch}{1.1}
\begin{tabular}{lccccccccc}
\hline
\hline

\multicolumn{9}{c}{\textbf{Collinear}}\\
\hline

\multicolumn{1}{m{2.2cm}}{\textbf{Fe Atom \#}}
& \multicolumn{1}{m{0.65cm}}{\centering \textbf{1}}
& \multicolumn{1}{m{0.65cm}}{\centering \textbf{2}}
& \multicolumn{1}{m{0.65cm}}{\centering \textbf{3}}
& \multicolumn{1}{m{0.65cm}}{\centering \textbf{4}}
& \multicolumn{1}{m{0.65cm}}{\centering \textbf{5}}
& \multicolumn{1}{m{0.65cm}}{\centering \textbf{6}}
& \multicolumn{1}{m{0.65cm}}{\centering \textbf{7}}
& \multicolumn{1}{m{0.65cm}}{\centering \textbf{8}}\\
\hline
\hline
Initial Moment & $\uparrow$ & $\downarrow$ & $\downarrow$ & $\uparrow$ & $\downarrow$ & $\uparrow$ & $\uparrow$ & $\downarrow$\\

\hline
Pure & 1.9 & -1.9 & -1.9 & 1.9 & -1.9 & 1.9 & 1.9 & -1.9\\

\hline
O$_{int A}$ & 1.9 & -1.8 & -1.7 & 1.8 & -1.8 & 1.8 & 1.8 & -1.8\\

\hline
O$_{int B}$ & 1.8 & -1.9 & -1.8 & 1.7 & -1.8 & 1.8 & 1.8 & -1.8\\

\hline
O$_{int A,B}$ & 2.3 & -2.3 & -1.1 & 1.1 & -1.1 & 1.1 & 1.8 & -1.8\\

\hline
O$_{int A,B, C}$ & \textbf{-2.5} & \textbf{2.8} & -1.7 & 1.5 & -1.6 & 1.5 & 2.0 & -2.1\\

\hline
Fe$_{vac.6}$ & 2.2 & -1.9 & -2.0 & 2.1 & -2.1 & - & 1.8 & -1.8\\

\hline
Fe$_{vac.1,6}$ & - & -1.3 & -1.3 & 2.2 & -1.3 & - & 2.2 & -1.3\\

\hline
Se$_{sub.doped(Te_{15})}$ & 1.9 & -1.9 & -1.9 & 1.9 & -1.9 & 1.9 & 1.9 & -1.9 \\

\hline
\hline
\end{tabular}
\end{center}
\caption{Calculated collinear magnetic moments of Fe atoms for
different configurations. Atom labels are as in
Fig.~\ref{fig:dos} and initial magnetic moment direction changes are highlighted in bold face.} \label{table1}
\end{table}

\begin{table}[t]
\begin{center}
\renewcommand{\arraystretch}{1.1}
\begin{tabular}{lccccccccc}
\hline
\hline

\multicolumn{9}{c}{\textbf{Bicollinear}}\\
\hline

\multicolumn{1}{m{2.2cm}}{\textbf{Fe Atom \#}}
& \multicolumn{1}{m{0.65cm}}{\centering \textbf{1}}
& \multicolumn{1}{m{0.65cm}}{\centering \textbf{2}}
& \multicolumn{1}{m{0.65cm}}{\centering \textbf{3}}
& \multicolumn{1}{m{0.65cm}}{\centering \textbf{4}}
& \multicolumn{1}{m{0.65cm}}{\centering \textbf{5}}
& \multicolumn{1}{m{0.65cm}}{\centering \textbf{6}}
& \multicolumn{1}{m{0.65cm}}{\centering \textbf{7}}
& \multicolumn{1}{m{0.65cm}}{\centering \textbf{8}}\\
\hline
\hline
Initial Moment & $\uparrow$ & $\uparrow$ & $\downarrow$ & $\downarrow$ & $\uparrow$ & $\uparrow$ & $\downarrow$ & $\downarrow$\\

\hline
Pure & 2.1 & 2.1 & -2.1 & -2.1 & 2.1 & 2.1 & -2.1 & -2.1\\

\hline
O$_{int A}$ & 2.3 & 1.6 & -1.4 & -2.0 & 1.0 & 1.6 & \textbf{1.0} & -2.0\\

\hline
O$_{int B}$ & 1.6 & 2.3 & -2.0 & -1.4 & 1.6 & 1.0 & -2.0 & \textbf{1.0}\\

\hline
O$_{int A,B}$ & 2.5 & 2.5 & -1.0 & -1.0 & 1.5 & 1.5 & -1.2 & -1.2\\

\hline
O$_{int A,B, C}$ & \textbf{-1.8} & 2.6 & -1.6 & \textbf{1.6} & 1.5 & 1.2 & -1.9 & -1.9 \\

\hline
Fe$_{vac.6}$ & 2.2 & \textbf{-1.7} & -1.5 & -0.5 & 2.2 & - & -1.5 & -1.6\\

\hline
Fe$_{vac.1,6}$ & - & 1.9 & -1.5 & -2.1 & 1.9 & - & -2.1 & -1.5\\

\hline
Se$_{sub.doped(Te_{15})}$ &  2.1 & 2.1 & -1.6 & \textbf{2.1} & 2.1 & -1.5 & -1.6 & -1.8\\

\hline
\hline
\end{tabular}
\end{center}
\caption{Calculated bicollinear magnetic moments of Fe atoms for
different configurations. Atom labels are as in
Fig.~\ref{fig:dos}.} \label{table2}
\end{table}

 We believe that, although the true, static antiferromagnetism vanishes with the above modulations, local magnetic order plays a crucial role
 at least as a precursor to superconductivity. There is new evidence for such behavior seen even in the cuprates.
 Recent RIXS (resonant inelastic X-ray scattering) experiments on the 123 cuprate have revealed an intense peak around 1.7 eV energy loss that is due to
 an optically forbidden $d-d$ transitions of unpaired holes from Cu$2^+$ to other d-orbitals (Ref.~\cite{tacon}).
 The significance of this work is that in overdoped, underdoped
 and superconducting samples, magnetic excitations are seen around the same energy region.

\subsection{Total energies}

 A comparison of  variationally evaluated total energies is one of the most useful and reliable outputs of a DFT-based calculation.
  Various modulation-induced structures were selected since, experimentally, these were identified as necessary precursors for  FeTe-derived superconductivity.
 Total energies shown in Fig.~\ref{fig:tot_energy}, for all the modulated structures show a very clear trend; i.e., the collinear-derived magnetic structure
becomes the more favorable one prior to possible superconducting transitions.

Here is a brief summary for the various structures:
(a) Pure Fe$_8$Te$_8$:  Collinear structure is more stable with Fe magnetic moments about 2$\mu_B$ per atom. (b) Fe$_{9}$Te$_8$: Bicollinear is more stable
with the extra Fe atom carrying a moment of about 2.5$\mu_B$. (c) Oxygen interstitial: Inserting one oxygen atom either at the interstitial site A or B
brings the total energies of the two magnetic structures close to one another and two inserted oxygens make it even closer. With three inserted oxygens,
the collinear structure clearly becomes the more favorable.  (d) Introduction of a single vacancy into the supercell induces a clear separation of
the total energies with collinear one being more stable; nevertheless, a second vacancy brings their energies closer together. (e) Se doping
 makes the collinear structure more stable.

  While a one particle mean-field theory alone is unlikely to explain pairing related phenomena in moderately correlated systems,
 the trends seen with oxygen insertions, Fe vacancies and
 Se doping appear  to confirm
several experimental findings, such as short-range magnetic order~\cite{wen}. The true, static AFM state, with large Fe moments,
 is lost due to oxygen insertions and other doping while some remnants of
 this lost state can be gleaned from the band calculations.
However, without further assumptions and work, these results alone are unable to demonstrate a pairing scenario as to where the holes
 are created and their role  due to these modulations.
 This is not surprising since
it is well known that the local approximations to exchange-correlation and mean-field effects usually wipe them out.
In addition,  Fermi surface topology-based nesting mechanisms have become popular and
 have been used to obtain various pairing mechanisms. However, in view of the loss of long range antiferromagnetic order and
 the onset of short range incommensurate order observed experimentally (Ref.~\cite{wen}) and  suggested by our supercell calculations, 
  we present a different pairing scenario, based on a spatially local mechanism, in the following section.

\begin{figure} 
\begin{center}
\includegraphics*[width=23pc]{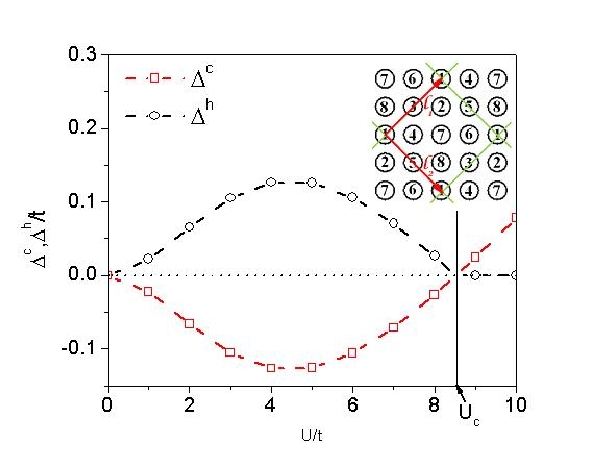}
\end{center}
\caption{Negative charge gap and positive spin gap in the 8-site Betts-cluster-based lattice.
 The 8-site Betts cell, which is periodically repeated with edge vectors (2,2) and (2,-2), is also shown here.}
 \label{fig:Betts_gap}
\end{figure}

\subsection{Charge and spin Pairing}

 Here we introduce a many-body, Hubbard Hamiltonian where
explicit fluctuations of paired holes can be systematically studied in a grand canonical ensemble~\cite{GWF, ANK}.
 Such calculations on a 8-site Betts cluster-based lattice (see Fig.~\ref{fig:Betts_gap})
 show that there are instabilities/fluctuations in charge and spin degrees of freedom under ``suitable" conditions.
 These fluctuations exist between a background AFM/SDW state and a two hole- or electron-doped state
 in a phase region identified as having a negative charge gap favoring charge and spin pairing (at a critical value of doping).
 Although our recent work~\cite{fang2}  refers to a single orbital model, we infer that it can be extended to two or multi-orbitals and still retain
 the negative charge gap, as long as hopping strength among similar orbitals is greater than that for different ones.

\begin{figure} 
\begin{center}
\includegraphics*[width=15pc]{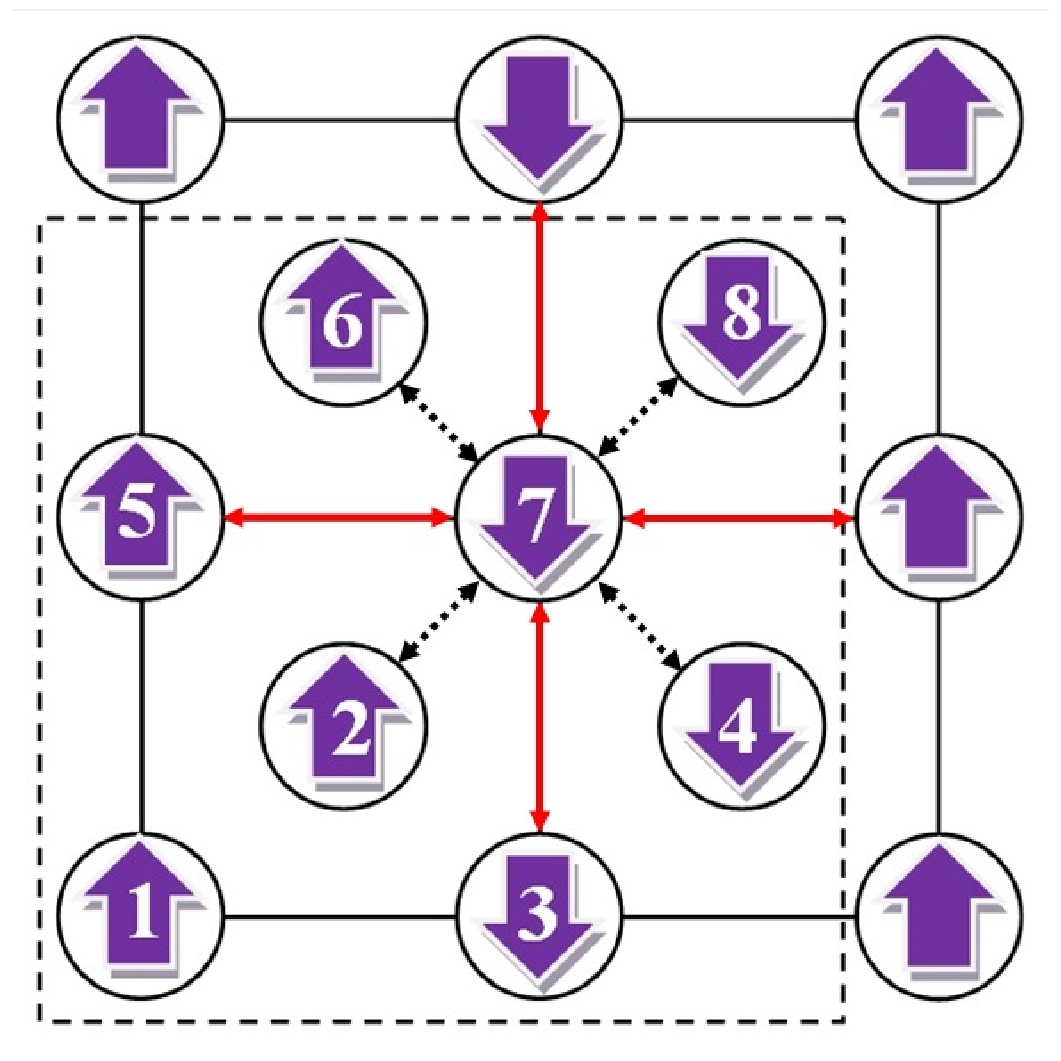}
\includegraphics*[width=15pc]{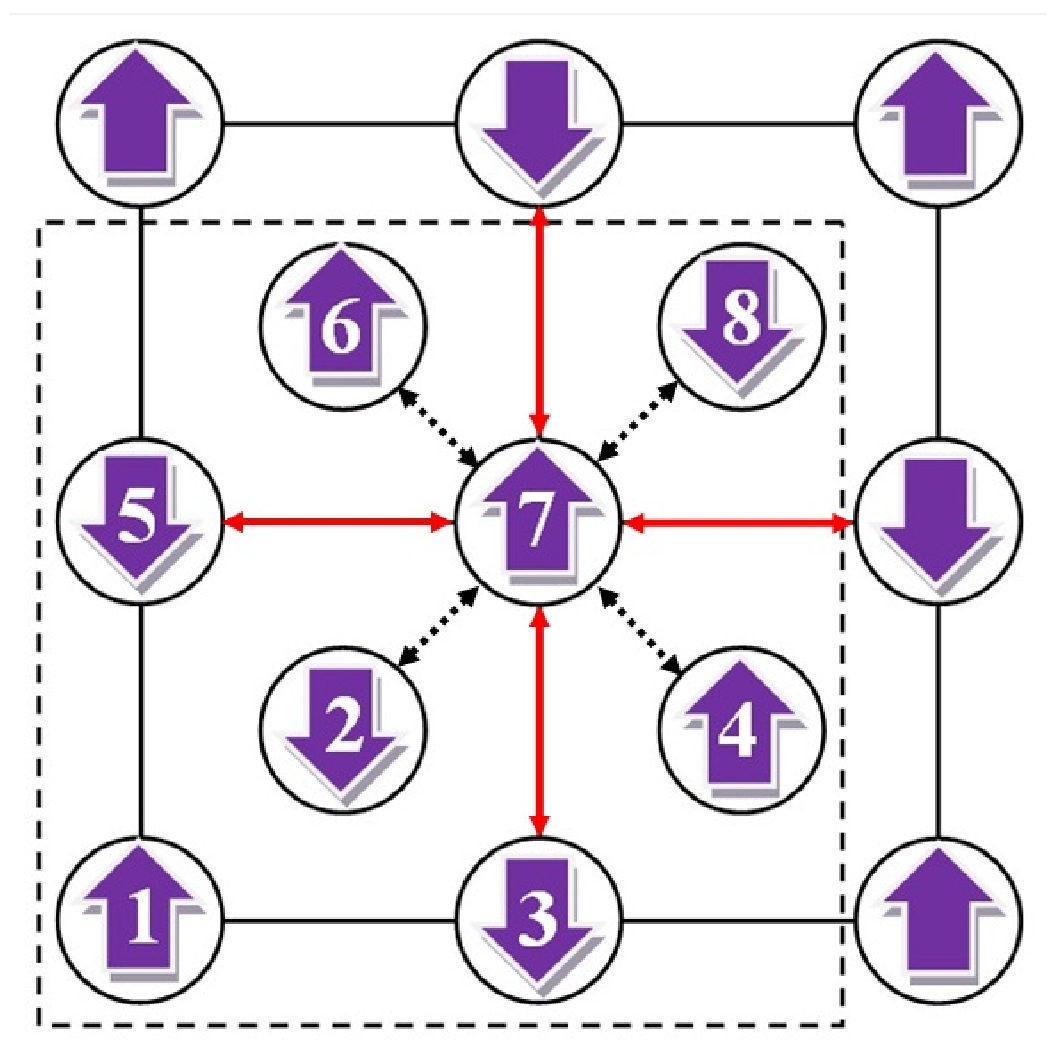}
\end{center}
\caption{Bicollinear (Top)  and collinear  (bottom) magnetic structures in the plane containing Fe atoms. Arrows identify nearest and next nearest neighbors
 of the seventh atom (as marked)  in the plane.}
\label{fig:moments}
\end{figure}

 Depending on the value of the electron-electron Coulomb repulsion $U$, there are certain regions in a 3-dimensional
 T(temperature)-$\mu$(chemical potential)-U phase space where
quantum critical points (QCPs) exist and drive  these fluctuations. For the 8-site Betts cluster-based lattice, with periodic boundary conditions,
 such a QCP has been found at $U=U_c=8.54$ (in units of $t$, the hopping parameter in the Hubbard model),
 so that a charge instability
exists in the region $0<U <U_c$ at suitable (doping) $\mu=\mu_c$ and T. This charge instability leads to fluctuations
 between a SDW/AFM state and a state with extra holes of
 paired (and unlike) spins at low temperature.
Note that even for very small $U$ (i.e., weakly correlated systems),  a negative charge gap and a coherent (positive) spin gap can exist.
Our grand conical ensemble-based studies show that these paired spins are stabilized by the (positive) spin gap up to a temperature $T_s$.
Clearly, with extra holes, it is easier for the charge carriers to move between sites, as seen in the oxygen-doped case.

An important point here is that such a scenario favors a background of collinear-derived states discussed in the previous sections,
 rather than the bicollinear one.  This is simply due to the differences in the neighboring spins which begin
 to appear at the  next nearest neighbor (nnn) level of the two structures (see Fig.~\ref{fig:moments}).
 For an electron (or a hole) to hop through the lattice and form a pair,
existence of (as many) unlike-spins as next nearest neighbors would be an asset (since like-spin neighbors would prevent hopping to that orbital)
provided that $U$ values are relatively small.
 This is what is seen in the collinear structure with 4 unlike-nnn-spins,
compared to 2 unlike-nnn-spins in the bicollinear structure (while at the nearest neighbor level, there is no difference).
 Our calculated charge gap for the 8-site Betts-cluster-based lattice shows that the
next nearest neighbor hopping, under certain conditions, can play a crucial, helpful role in  the charge and spin pairing instability~\cite{fang2}.

\subsection{Summary}

  Our band calculations, which are consistent with several experimental results, are used  to identify states that act as precursors to superconductivity.
In every case considered for the Fe-chalcogenide under consideration, a collinear-derived, damped SDW is predicted.
A theory based on the Hubbard model,  with a weak onsite Coulomb repulsion, is able to explain possible charge and
 spin fluctuations starting from these precursors.
We believe that this work lays the ground work for a common understanding of superconductivity in the chalcogenides, pnictides and cuprates.

\subsection{Acknowledgments}
We thank Profs. J. I. Budnick and B. O. Wells for useful
discussions. The authors also acknowledge the computing facilities
provided by the Center for Functional Nanomaterials,
 Brookhaven National Laboratory, which is supported by the U.S. Department of Energy, Office of Basic Energy Sciences, under Contract No. DE-AC02-98CH10886.

\end{document}